\documentclass[preprint,aps,prd,nofootinbib,superscriptaddress,floatfix]{revtex4-1}
\usepackage{amsmath,amssymb}
\usepackage{bm}
\usepackage{graphicx}
\makeatletter
\def\label#1{\@bsphack
  \begingroup
  \UseHookWithArguments{label}{1}{#1}
  \protected@write\@auxout{}
         {\string\newlabel{#1}{{\@currentlabel}{\thepage}
          {\@currentlabelname}{\@currentHref}{\@kernel@reserved@label@data}}}
  \endgroup
  \@esphack}
\makeatother
\usepackage{hyperref}

\renewcommand{\arraystretch}{1.14}
\graphicspath{{figures/}}

\newcommand{\dd}{\mathrm{d}}
\newcommand{\eeds}{Einstein--Euler--Heisenberg--de Sitter}

\begin{document}

\title{Scattering and Hawking  Radiation from Einstein--Euler--Heisenberg--de Sitter Black Holes}

\author{Jayden Tan\\Equilibrium Research Unit, Analysis of Systemic Complexity\\\#02-01, 99 Portsdown Road, Singapore 139299}
\date{jaydtan2002@proton.me}

\date{\today}

\begin{abstract}
We compute neutral scalar and neutral massless Dirac greybody factors and Hawking spectra for the positive-cosmological-constant branch of the Einstein--Euler--Heisenberg black hole. The finite region between the black-hole and cosmological horizons is treated as a two-sided scattering problem, with direct numerical integrations providing the transmission coefficients and a sixth-order WKB calculation used as a local check near the barrier top. Along the reference family studied here, increasing the Euler--Heisenberg coupling raises the dominant scalar and Dirac barriers and shifts the half-transmission frequencies upward. At fixed charge and nonlinear coupling, increasing the cosmological constant contracts the static patch and lowers the dominant greybody thresholds. The luminosity is controlled as much by the de Sitter temperature prescription as by the greybody factors: event-horizon prescriptions brighten the emission as the nonlinear correction grows, whereas effective static-patch temperatures give much smaller rates and can reverse the trend. Thus the evaporation interpretation is not unique unless the temperature convention is specified explicitly.
\end{abstract}

\maketitle

\section{Introduction}
\label{sec:introduction}

Black holes are at once thermodynamic objects and scattering systems. Hawking's result tells us that horizons populate quantum fields thermally, but the radiation received away from the horizon is not the spectrum of an ideal furnace. Before escaping, each wave must cross the curved exterior of the black hole. The frequency-dependent transmission probabilities produced by this crossing are the greybody factors, and they determine how the near-horizon thermal bath is reshaped into the flux measured in the static region or at infinity \cite{Hawking1975,Unruh1976}.

This filtering is interesting for several reasons. Greybody factors encode absorption probabilities, partial-wave cross sections and the relative weight of different spin channels in Hawking emission. They are therefore sensitive to details of the geometry that may be almost invisible in a purely thermodynamic treatment. A deformation that makes a black hole hotter need not make it brighter if the exterior barrier grows at the same time; conversely, a more transparent barrier can be overwhelmed by a colder horizon. The luminosity is the result of this competition between horizon temperature and wave propagation.

Greybody factors also sit close to quasinormal modes. The same radial master equation describes both problems; only the boundary conditions are changed. Real-frequency scattering asks how much of an incoming wave crosses the barrier, while quasinormal ringing asks for the complex frequencies that satisfy purely horizon boundary conditions. In this sense, greybody factors and quasinormal modes are two faces of the same wave dynamics: one records the transmission of forced waves, the other the natural damped response of the black hole. This connection is especially visible in WKB treatments, where the height and curvature of the potential barrier control both the onset of transmission and the quasinormal spectrum. 

The de Sitter case adds a further layer. The exterior is no longer an infinite scattering line but a finite chamber bounded by the black-hole and cosmological horizons. There is no spatial infinity at which to normalize the static Killing vector, and the temperature assigned to the black-hole horizon is therefore not unique. Surface-gravity, Bousso--Hawking and effective static-patch prescriptions may lead to different emission rates even when the greybody factors themselves are fixed. Perturbations, classical (quasinormal) and quantum (Hawking) spectra of asymptotically de Sitter black holes were studied in a great number of publications \cite{Jansen:2017oag,Konoplya:2004uk,Aragon:2020qdc,Zhidenko:2003wq,Dyatlov:2010hq,Kanti:2005xa,Mo:2018nnu,Jing:2003wq,Konoplya:2008au,Cuyubamba:2016cug,Molina:2003ff,Konoplya:2007jv,Lutfuoglu:2026pgn,Lutfuoglu:2025eik,Konoplya:2013sba,Bolokhov:2026dfg,Bolokhov:2024ixe,Skvortsova:2023zmj,Konoplya:2017ymp,Skvortsova:2026idf,Malik:2026oxx,Malik:2025ava}.

Black holes in nonlinear electrodynamics provide a particularly clear stage on which such issues can be examined. The Euler--Heisenberg effective theory is the low-energy one-loop description of quantum-electrodynamic vacuum polarization in slowly varying electromagnetic fields \cite{HeisenbergEuler1936}. Its quartic field-strength terms are faint in ordinary weak-field environments, yet near a charged black hole they become a controlled way to deform the geometry, its thermodynamics and the propagation of fields. Quasinormal modes of black holes as well as WKB grey-body factors in this theory have been recently considered in \cite{Lutfuoglu:2025bsf,Malik:2025erb}.

The particular Einstein--Euler--Heisenberg metric used here has already been studied from several complementary viewpoints. Magos and Bret\'on introduced the cosmological constant and analyzed the anti--de Sitter branch as a black-hole-chemistry system, including the equation of state, critical behavior and Gibbs free energy \cite{MagosBreton2020}. Subsequent work focused mainly on thermodynamic properties of the same AdS branch: Ye et al. studied how the Euler--Heisenberg coupling changes phase transitions and Ruppeiner geometry in the extended phase space \cite{YeChenLiWei2022}; Dai, Zhao and Zhang treated the first-order transition as a stochastic process on the free-energy landscape \cite{DaiZhaoZhang2023}; Gogoi and Phukon examined the ensemble dependence of the thermodynamic topology \cite{GogoiPhukon2024}; and Zhao and Cheng compared canonical and grand-canonical stability, heat capacity and phase structure \cite{ZhaoCheng2024}. Optical properties have also been investigated for the same AdS/dS metric: Jafarzade et al. derived the nonlinear-electrodynamic effective metric for photons and used it to study photon trajectories, shadows, emission rates and light deflection against Event Horizon Telescope constraints \cite{JafarzadeBazyarSaghafiNozari2025}. These studies establish the thermodynamic and optical relevance of the background, but they do not provide neutral scalar and Dirac greybody factors in the finite de Sitter static patch.

In this work we study the positive-cosmological-constant branch of the Einstein--Euler--Heisenberg solution, compute neutral scalar and Dirac greybody factors, compare direct integration with WKB estimates near the barrier top, and follow how Hawking spectra respond to the available de Sitter temperature prescriptions.

The paper is organized as follows. Section~\ref{sec:ehds_geometry} records the background geometry and horizon normalization. Section~\ref{sec:temperature_conventions} reviews the temperature prescriptions needed in a two-horizon static patch. Sections~\ref{sec:wave_equations}--\ref{sec:greybody_results} develop the perturbation equations, numerical method and greybody results, while Sec.~\ref{sec:emission_rates} applies the resulting transmission coefficients to Hawking emission.

\section{The Einstein--Euler--Heisenberg--de Sitter geometry}
\label{sec:ehds_geometry}

Static black-hole solutions in Einstein--Euler--Heisenberg theory were constructed first without a cosmological constant \cite{YajimaTamaki2001}. The solution relevant for the present work is the extension by Magos and Bret\'on, who introduced a cosmological constant into the Einstein--Euler--Heisenberg system and obtained a four-parameter family characterized by mass $M$, electric charge $Q$, cosmological constant $\Lambda$ and Euler--Heisenberg coupling $a$ \cite{MagosBreton2020}. Although that paper is usually discussed in the anti--de Sitter thermodynamic context, the metric itself admits either sign of $\Lambda$. The black hole studied here is the positive-cosmological-constant branch, $\Lambda>0$, which we refer to as the \eeds{} black hole. Recent optical studies of the same Einstein--Euler--Heisenberg AdS/dS family use the same metric as their background geometry \cite{JafarzadeBazyarSaghafiNozari2025}.

We work in geometrized units $G=c=\hbar=k_B=1$ and use the mostly-plus signature. The action can be written as
\begin{equation}
\label{eq:ehds_action}
 S=\frac{1}{4\pi}\int \dd^4x\sqrt{-g}
 \left[\frac{1}{4}\left(R-2\Lambda\right)-\mathcal{L}(\mathcal{F},\mathcal{G})\right],
\end{equation}
where
\begin{equation}
\label{eq:eh_lagrangian}
 \mathcal{L}(\mathcal{F},\mathcal{G})
 =-\mathcal{F}+\frac{a}{2}\mathcal{F}^2+\frac{7a}{8}\mathcal{G}^2,
 \qquad
 \mathcal{F}=\frac{1}{4}F_{\mu\nu}F^{\mu\nu},
 \qquad
 \mathcal{G}=\frac{1}{4}F_{\mu\nu}{^*F}^{\mu\nu}.
\end{equation}
Here $S$ is the total action, $g$ is the determinant of the spacetime metric, $R$ is the Ricci scalar, $\Lambda$ is the cosmological constant, $\mathcal{L}$ is the nonlinear electromagnetic Lagrangian, and $\mathcal{F}$ and $\mathcal{G}$ are the two electromagnetic invariants.
Here $F_{\mu\nu}$ is the electromagnetic field strength and ${^*F}^{\mu\nu}$ is its dual. The parameter $a$ measures the strength of the leading Euler--Heisenberg correction; setting $a=0$ removes the nonlinear electrodynamic contribution and restores the Maxwell limit.

The static, spherically symmetric line element is
\begin{equation}
\label{eq:ehds_metric}
 \dd s^2=-f(r)\,\dd t^2+\frac{\dd r^2}{f(r)}+r^2\dd\Omega_2^2,
\end{equation}
with lapse function
\begin{equation}
\label{eq:ehds_lapse}
 f(r)=1-\frac{2M}{r}+\frac{Q^2}{r^2}-\frac{\Lambda r^2}{3}
      -\frac{aQ^4}{20r^6}.
\end{equation}
Here $t$ is the static time coordinate, $r$ is the areal radius, $\dd\Omega_2^2$ is the metric on the unit two-sphere, $f(r)$ is the lapse function, $M$ is the mass parameter, and $Q$ is the electric charge.
The last term is the Euler--Heisenberg correction. The standard limits are immediate: $a\to0$ gives Reissner--Nordstr\"om--de Sitter, $\Lambda\to0$ gives the asymptotically flat Einstein--Euler--Heisenberg black hole, and $Q\to0$ gives Schwarzschild--de Sitter. For positive physical Euler--Heisenberg coupling $a$, the short-distance behavior is dominated by $f(r)\sim-aQ^4/(20r^6)$, so this solution should not be confused with a regular black hole; the nonlinear electromagnetic correction changes the central singularity rather than removing it.

For a purely electric configuration one may choose a gauge potential $A=-\Phi(r)\dd t$, with
\begin{equation}
\label{eq:ehds_potential_field}
 \Phi(r)=\frac{Q}{r}\left(1-\frac{aQ^2}{10r^4}\right),
 \qquad
 \mathcal{E}(r)=-\Phi'(r)=\frac{Q}{r^2}\left(1-\frac{aQ^2}{2r^4}\right).
\end{equation}
Here $A$ is the electromagnetic one-form, $\Phi(r)$ is the electrostatic potential, $\mathcal{E}(r)$ is the radial electric field, and a prime denotes differentiation with respect to $r$.
The Maxwell expressions are recovered when $a=0$. In de Sitter space the electrostatic potential relevant for thermodynamics or charged-field boundary conditions is a potential difference between two chosen horizons or between a horizon and a selected reference worldline, rather than a potential measured at spatial infinity.

The horizons are the positive real roots of $f(r)=0$. Multiplying by $r^6$ gives the polynomial form
\begin{equation}
\label{eq:ehds_horizon_equation}
 \frac{\Lambda}{3}r_h^8-r_h^6+2Mr_h^5-Q^2r_h^4+\frac{aQ^4}{20}=0.
\end{equation}
Here $r_h$ denotes a generic positive horizon radius obtained from $f(r_h)=0$.
The positive roots can include an inner Cauchy horizon $r_-$ in addition to the event horizon $r_b$ and the cosmological horizon $r_c$. When all three are present they are ordered as
\begin{equation}
\label{eq:ehds_horizon_order}
 0<r_-<r_b<r_c.
\end{equation}
For the scattering problem only the outer static patch is needed, and this region is the finite interval $r_b<r<r_c$. Degenerate limits occur when two neighboring roots merge. A double root $r_h$ obeys both $f(r_h)=0$ and $f'(r_h)=0$, equivalently
\begin{equation}
\label{eq:ehds_degenerate_condition}
 1-\frac{Q^2}{r_h^2}+\frac{aQ^4}{4r_h^6}-\Lambda r_h^2=0.
\end{equation}
Depending on which roots coincide, this condition describes a cold black hole with $r_-=r_b$ or a Nariai-type limit with $r_b=r_c$.

It is useful to solve the event-horizon condition for the mass parameter. Taking $r_b$ as the black-hole event-horizon radius gives
\begin{equation}
\label{eq:ehds_mass_horizon}
 M=\frac{r_b}{2}\left(1+\frac{Q^2}{r_b^2}-\frac{\Lambda r_b^2}{3}
 -\frac{aQ^4}{20r_b^6}\right).
\end{equation}
This representation is convenient because it fixes the black-hole horizon and leaves the charge, cosmological constant and Euler--Heisenberg coupling as dimensionless deformation parameters. Defining
\begin{equation}
\label{eq:ehds_dimensionless_parameters}
 x=\frac{r}{r_b},
 \qquad
 q=\frac{Q}{r_b},
 \qquad
 \alpha=\frac{a}{r_b^2},
 \qquad
 \lambda=\Lambda r_b^2,
\end{equation}
where $x$ is the dimensionless radius, $q$ is the dimensionless charge, $\alpha$ is the dimensionless Euler--Heisenberg coupling, and $\lambda$ is the dimensionless cosmological constant.
The normalized lapse becomes
\begin{equation}
\label{eq:ehds_normalized_lapse}
 f(x)=1-\frac{1+q^2-\lambda/3-\alpha q^4/20}{x}
 +\frac{q^2}{x^2}-\frac{\lambda x^2}{3}-\frac{\alpha q^4}{20x^6},
 \qquad f(1)=0.
\end{equation}
The cosmological horizon is then the largest root $x_c>1$ of Eq.~\eqref{eq:ehds_normalized_lapse}. This form will be used later to define dimensionless tortoise coordinates and frequency variables for the scattering problem.

The surface gravity associated with a simple horizon $r_h$ is $\kappa_h=f'(r_h)/2$, up to the sign set by the horizon generator. With the coordinate normalization used so far, the black-hole event-horizon temperature is
\begin{equation}
\label{eq:ehds_black_hole_temperature}
 T_b=\frac{f'(r_b)}{4\pi}
 =\frac{1}{4\pi r_b}\left(1-\frac{Q^2}{r_b^2}+\frac{aQ^4}{4r_b^6}-\Lambda r_b^2\right),
\end{equation}
or, in the normalized variables,
\begin{equation}
\label{eq:ehds_dimensionless_temperature}
 r_bT_b=\frac{1-q^2+\alpha q^4/4-\lambda}{4\pi}.
\end{equation}
Here $T_b$ is the event-horizon temperature associated with the coordinate normalization of Eq.~\eqref{eq:ehds_metric}.
For the cosmological horizon one uses $T_c=-f'(r_c)/(4\pi)$ when $f'(r_c)<0$. Except for special lukewarm configurations, $T_b$ and $T_c$ are different, so the de Sitter black hole is not in global thermal equilibrium. Section~\ref{sec:temperature_conventions} summarizes the main temperature conventions that may be used in such a multi-horizon spacetime. The Bekenstein--Hawking entropy of the event horizon remains
\begin{equation}
\label{eq:ehds_entropy}
 S_b=\frac{A_b}{4}=\pi r_b^2,
\end{equation}
Here $S_b$ is the event-horizon entropy and $A_b=4\pi r_b^2$ is the event-horizon area. The area law is unchanged because the gravitational sector is the Einstein--Hilbert one; the nonlinear electrodynamics changes the matter stress tensor and the mass--horizon relation, not the entropy functional.

Finally, one should distinguish the background metric in Eq.~\eqref{eq:ehds_metric} from the effective optical metric that can govern high-frequency electromagnetic perturbations in nonlinear electrodynamics. Neutral scalar, Dirac and gravitational test fields couple directly to the spacetime metric, whereas electromagnetic waves can experience an effective geometry depending on derivatives of the nonlinear Lagrangian. The calculations below use Eq.~\eqref{eq:ehds_lapse} as the background geometry and treat the finite de Sitter interval $r_b<r<r_c$ as the natural radial domain for neutral scalar and Dirac greybody factors and Hawking spectra.

\section{Temperature conventions for the de Sitter branch}
\label{sec:temperature_conventions}

In asymptotically flat black-hole spacetimes the normalization of the static Killing vector is fixed at spatial infinity, and the Hawking temperature is usually identified unambiguously with the surface gravity divided by $2\pi$. A de Sitter black hole has no spatial infinity and possesses both a black-hole horizon and a cosmological horizon. Consequently there is not a unique operational temperature for the whole static patch. This ambiguity is not a pathology of the \eeds{} geometry; it is the same issue discussed for Schwarzschild--de Sitter emission by Pappas and Kanti, who compared bare horizon temperatures, the Bousso--Hawking normalized temperature and several effective temperatures in the Hawking spectrum \cite{KantiPappas2017Effective,PappasKanti2017Role}. We therefore keep the temperature convention explicit throughout the paper.

The first and most local convention is the surface-gravity, or ``bare'', temperature. For a simple horizon $r_h$ one defines
\begin{equation}
\label{eq:ehds_general_surface_temperature}
 T_h^{\rm sg}=\frac{|f'(r_h)|}{4\pi}.
\end{equation}
The subscript $h$ labels the chosen horizon and the superscript ``sg'' denotes the unnormalized surface-gravity temperature.
Using the horizon equation to eliminate $M$, this can be written for any horizon as
\begin{equation}
\label{eq:ehds_surface_derivative_horizon}
 f'(r_h)=\frac{1}{r_h}-\frac{Q^2}{r_h^3}+\frac{aQ^4}{4r_h^7}-\Lambda r_h.
\end{equation}
Thus the event-horizon and cosmological-horizon temperatures are
\begin{align}
\label{eq:ehds_bare_temperatures}
 T_b^{\rm sg}
 &=\frac{1}{4\pi r_b}\left(1-\frac{Q^2}{r_b^2}+\frac{aQ^4}{4r_b^6}-\Lambda r_b^2\right),\\
 T_c^{\rm sg}
 &=\frac{1}{4\pi r_c}\left(\Lambda r_c^2-1+\frac{Q^2}{r_c^2}-\frac{aQ^4}{4r_c^6}\right).
\end{align}
The absolute value in Eq.~\eqref{eq:ehds_general_surface_temperature} is responsible for the sign change between the two horizons. In the normalized variables of Eq.~\eqref{eq:ehds_dimensionless_parameters},
\begin{equation}
\label{eq:ehds_dimensionless_cosmological_temperature}
 r_bT_c^{\rm sg}=\frac{1}{4\pi x_c}\left(\lambda x_c^2-1+\frac{q^2}{x_c^2}-\frac{\alpha q^4}{4x_c^6}\right),
\end{equation}
where $x_c=r_c/r_b$ is the largest positive root of Eq.~\eqref{eq:ehds_normalized_lapse}.

A second convention normalizes the static Killing vector at the radius where a static observer has zero four-acceleration. This is the Bousso--Hawking prescription \cite{BoussoHawking1996}, often used in de Sitter black-hole radiation calculations. Let
\begin{equation}
\label{eq:ehds_zero_acceleration_radius}
 r_0\in(r_b,r_c),\qquad f'(r_0)=0,
\end{equation}
be this geodesic static radius. Equivalently,
\begin{equation}
\label{eq:ehds_zero_acceleration_equation}
 \frac{M}{r_0^2}-\frac{Q^2}{r_0^3}-\frac{\Lambda r_0}{3}
 +\frac{3aQ^4}{20r_0^7}=0.
\end{equation}
The normalized Killing vector is then $K=\gamma_t\partial_t$ with
\begin{equation}
\label{eq:ehds_killing_normalization}
 \gamma_t=\frac{1}{\sqrt{f(r_0)}}.
\end{equation}
Here $r_0$ is the geodesic static radius, $K$ is the normalized timelike Killing vector, and $\gamma_t$ is the corresponding normalization factor.
The corresponding normalized horizon temperatures are
\begin{equation}
\label{eq:ehds_bousso_hawking_temperatures}
 T_b^{\rm norm}=\frac{T_b^{\rm sg}}{\sqrt{f(r_0)}},
 \qquad
 T_c^{\rm norm}=\frac{T_c^{\rm sg}}{\sqrt{f(r_0)}}.
\end{equation}
In dimensionless form $x_0=r_0/r_b$ is fixed by
\begin{equation}
\label{eq:ehds_dimensionless_zero_acceleration}
 \frac{1+q^2-\lambda/3-\alpha q^4/20}{x_0^2}
 -\frac{2q^2}{x_0^3}-\frac{2\lambda x_0}{3}
 +\frac{3\alpha q^4}{10x_0^7}=0,
\end{equation}
with $1<x_0<x_c$. The event-horizon expression becomes
\begin{equation}
\label{eq:ehds_dimensionless_bousso_hawking_temperature}
 r_bT_b^{\rm norm}=\frac{1-q^2+\alpha q^4/4-\lambda}{4\pi\sqrt{f(x_0)}}.
\end{equation}
Here $x_0=r_0/r_b$ is the dimensionless zero-acceleration radius and $T_b^{\rm norm}$ is the Bousso--Hawking-normalized event-horizon temperature.
This normalization is especially important near Nariai-type limits: the bare surface-gravity temperature may vanish because the two horizons coalesce, while the normalization factor can also become singular, leaving a different limiting temperature.

A third class of prescriptions assigns a single effective temperature to the whole static patch. These definitions are useful when one treats the black-hole and cosmological horizons as parts of a combined thermodynamic system, but they are not equivalent to a local detector temperature at one horizon. Following the notation used in the Schwarzschild--de Sitter analysis of Pappas and Kanti \cite{KantiPappas2017Effective,PappasKanti2017Role}, two simple effective choices are
\begin{equation}
\label{eq:ehds_effective_temperatures_pm}
 T_{\rm eff}^{(-)}=\left(\frac{1}{T_c^{\rm sg}}-\frac{1}{T_b^{\rm sg}}\right)^{-1},
 \qquad
 T_{\rm eff}^{(+)}=\left(\frac{1}{T_c^{\rm sg}}+\frac{1}{T_b^{\rm sg}}\right)^{-1}.
\end{equation}
The superscripts $(-)$ and $(+)$ label the difference and harmonic effective-temperature prescriptions, respectively.
One may also mix the cosmological-horizon temperature with the normalized black-hole temperature,
\begin{equation}
\label{eq:ehds_effective_bousso_hawking_temperature}
 T_{\rm eff}^{\rm norm}=\left(\frac{1}{T_c^{\rm sg}}-\frac{1}{T_b^{\rm norm}}\right)^{-1},
\end{equation}
or use an effective first-law temperature. If the Schwarzschild--de Sitter expression is adopted as a phenomenological guide, it reads
\begin{equation}
\label{eq:ehds_effective_eiw_temperature}
 T_{\rm eff}^{\rm EIW}=\frac{r_b^4T_c^{\rm sg}+r_c^4T_b^{\rm sg}}
 {(r_b+r_c)(r_c^3-r_b^3)}.
\end{equation}
Here $T_{\rm eff}^{\rm norm}$ is the effective prescription built from $T_c^{\rm sg}$ and $T_b^{\rm norm}$, while $T_{\rm eff}^{\rm EIW}$ denotes the effective first-law-inspired temperature used as a phenomenological comparison.
For the charged \eeds{} black hole, however, Eq.~\eqref{eq:ehds_effective_eiw_temperature} should be regarded as a convention unless the corresponding effective first law is rederived with the charge $Q$ and Euler--Heisenberg coupling $a$ specified as fixed thermodynamic parameters or varied work terms.

Finally, a static detector at radius $r$ measures a Tolman-redshifted local temperature,
\begin{equation}
\label{eq:ehds_tolman_temperature}
 T_{{\rm loc},h}(r)=\frac{T_h^{\rm sg}}{\sqrt{f(r)}} ,
 \qquad r_b<r<r_c,
\end{equation}
Here $T_{{\rm loc},h}(r)$ is the temperature measured by a static detector at radius $r$ for radiation associated with horizon $h$. This quantity diverges as the detector approaches the corresponding horizon and is distinct from the coordinate temperature inserted into a flux formula at a chosen boundary. In the greybody and Hawking spectra below, we therefore state explicitly which of $T_b^{\rm sg}$, $T_b^{\rm norm}$ or an effective static-patch temperature is being used. Charged-particle emission would also require the appropriate electrostatic chemical potential; the present neutral test-field discussion needs only the thermal factor.

\section{Scalar and Dirac perturbations}
\label{sec:wave_equations}

The tortoise coordinate is defined by
\begin{equation}
\label{eq:tortoise}
 \frac{\dd r_*}{\dd r}=\frac{1}{f(r)} .
\end{equation}
Here $r_*$ is the tortoise coordinate that maps the finite interval $r_b<r<r_c$ to an infinite scattering line.
In the de Sitter branch $r_*\to-\infty$ at the black-hole event horizon and $r_*\to+\infty$ at the cosmological horizon. The static patch therefore provides a two-sided scattering problem, with the black-hole horizon on the left and the cosmological horizon on the right.

For a neutral massless scalar field, $\Box\Phi=0$, we separate variables as
\begin{equation}
\label{eq:scalar_separation}
 \Phi=\frac{\Psi_{\omega\ell}(r)}{r}Y_{\ell m}(\theta,\phi)e^{-i\omega t}.
\end{equation}
Here $\Phi$ is the scalar field, $\Psi_{\omega\ell}$ is the radial amplitude, $Y_{\ell m}$ is a spherical harmonic, $\omega$ is the real frequency, and $\ell,m$ are the scalar angular quantum numbers.
The radial equation is (see ~\cite{Carter:1968ks,Konoplya:2018arm})
\begin{equation}
\label{eq:scalar_master} \frac{\dd^2\Psi_{\omega\ell}}{\dd r_*^2}
 +\left[\omega^2-V_s(r)\right]\Psi_{\omega\ell}=0,
 \qquad
 V_s(r)=f(r)\left[\frac{\ell(\ell+1)}{r^2}+\frac{f'(r)}{r}\right].
\end{equation}
Here $V_s(r)$ is the scalar effective potential and the prime on $f$ denotes $\dd f/\dd r$.
The scalar multiplicity is $2\ell+1$.

For a neutral massless Dirac field the radial system can be written in a supersymmetric pair of Schr\"odinger-type equations (see, for instance \cite{Konoplya:2017tvu,Cho:2004wj,Kanti:2006ua,Cho:2003qe,Konoplya:2007zx}),
\begin{equation}
\label{eq:dirac_master}
 \frac{\dd^2 Z_\pm}{\dd r_*^2}+\left[\omega^2-V_\pm(r)\right]Z_\pm=0,
 \qquad
 V_\pm(r)=W^2\pm\frac{\dd W}{\dd r_*},
 \qquad
 W(r)=\frac{\kappa\sqrt{f(r)}}{r}.
\end{equation}
Here $\kappa=1,2,\ldots$ labels the angular Dirac modes. $Z_\pm$ are the two radial Dirac master functions, $V_\pm$ are the supersymmetric partner potentials, and $W(r)$ is the Dirac superpotential; the derivative $\dd W/\dd r_*$ is evaluated using $\dd/\dd r_*=f(r)\dd/\dd r$. The two partner potentials are isospectral for the transmission problem, so we use $V_+$ in the numerical calculations. For one four-component massless Dirac field the multiplicity of the $\kappa$ channel is $4\kappa$; a Majorana field would have half this degeneracy.

All numerical quantities below are reported in units of the event-horizon radius. The reference path through parameter space is
\begin{equation}
\label{eq:reference_point_family}
 q=\frac{Q}{r_b}=0.60,
 \qquad
 \lambda=\Lambda r_b^2=0.04,
 \qquad
 \alpha=\frac{a}{r_b^2}=0,5,10,15.
\end{equation}
This benchmark fixes the dimensionless charge $q$ and cosmological constant $\lambda$ while varying only the dimensionless nonlinear-electrodynamic coupling $\alpha$.
This choice keeps a broad static interval between the black-hole and cosmological horizons while making the Euler--Heisenberg correction visible. The corresponding horizon and temperature data are listed in Table~\ref{tab:ehds_parameters}.

\begin{table*}[t]
\centering
\caption{Reference \eeds{} backgrounds used in the greybody calculation. The event horizon is fixed at $r_b=1$, with $q=Q/r_b=0.60$, $\lambda=\Lambda r_b^2=0.04$ and $\alpha=a/r_b^2=0,5,10,15$. The columns give the mass parameter, the cosmological horizon $x_c=r_c/r_b$, the zero-acceleration radius $x_0=r_0/r_b$, and three representative temperatures in units of $r_b^{-1}$.}
\label{tab:ehds_parameters}
\begin{ruledtabular}
\begin{tabular}{ccccccc}
$\alpha$ & $M/r_b$ & $x_c$ & $x_0$ & $r_bT_b^{\rm sg}$ & $r_bT_b^{\rm norm}$ & $r_bT_c^{\rm sg}$ \\
$0$ & $0.673333$ & $7.916618$ & $3.497413$ & $4.775\times10^{-2}$ & $6.882\times10^{-2}$ & $1.521\times10^{-2}$ \\
$5$ & $0.657133$ & $7.937941$ & $3.462210$ & $6.064\times10^{-2}$ & $8.657\times10^{-2}$ & $1.530\times10^{-2}$ \\
$10$ & $0.640933$ & $7.959078$ & $3.426224$ & $7.353\times10^{-2}$ & $1.040\times10^{-1}$ & $1.539\times10^{-2}$ \\
$15$ & $0.624733$ & $7.980031$ & $3.389411$ & $8.642\times10^{-2}$ & $1.211\times10^{-1}$ & $1.549\times10^{-2}$ \\
\end{tabular}
\end{ruledtabular}
\end{table*}

\section{Numerical and WKB methods}
\label{sec:numerical_methods}

The transmission coefficient is computed by imposing a unit-amplitude wave at the black-hole horizon and reading off the incident amplitude at the cosmological horizon. With the time dependence $e^{-i\omega t}$, the boundary conditions are
\begin{align}
\label{eq:boundary_conditions}
 \Psi &\sim e^{-i\omega r_*}, \qquad r\to r_b,\\
 \Psi &\sim A_{\rm in}e^{-i\omega r_*}+A_{\rm out}e^{+i\omega r_*},
 \qquad r\to r_c.
\end{align}
The transmitted wave at the black-hole horizon is normalized to unit amplitude, while $A_{\rm in}$ and $A_{\rm out}$ are the incident and reflected amplitudes measured near the cosmological horizon.
The greybody factor is then
\begin{equation}
\label{eq:transmission_from_amplitude}
 \Gamma_{\ell,\kappa}(\omega)=\frac{1}{|A_{\rm in}|^2},
 \qquad
 \mathcal{R}=\left|\frac{A_{\rm out}}{A_{\rm in}}\right|^2,
 \qquad
 \Gamma+\mathcal{R}=1.
\end{equation}
Here $\Gamma_{\ell,\kappa}$ denotes the transmission probability for the relevant scalar or Dirac channel and $\mathcal{R}$ is the corresponding reflection probability.
In practice the radial equation is integrated over a dense nonuniform grid in $r$, clustered near both horizons, after constructing $r_*(r)$ by numerical quadrature of Eq.~\eqref{eq:tortoise}. The wave equation is solved as a first-order complex system in $r_*$. The amplitudes $A_{\rm in}$ and $A_{\rm out}$ are extracted from the last part of the numerical solution close to the cosmological horizon. The largest flux-balance error over all tabulated curves is $6.9\times10^{-5}$, which is small compared with the trends discussed below.

The direct numerical calculation used for the greybody factors can be summarized without reference to any implementation details as follows. For each choice of $(q,\lambda,\alpha)$ and each partial wave, the cosmological horizon $x_c=r_c/r_b$ is found as the largest root of Eq.~\eqref{eq:ehds_normalized_lapse}. The tortoise map is then constructed on the finite radial interval $1<x<x_c$ by integrating
\begin{equation}
\label{eq:numerical_tortoise_quadrature}
 r_*(x)=r_b\int_{1+\epsilon_b}^{x}\frac{\dd y}{f(y)},
 \qquad 1+\epsilon_b\leq x\leq x_c-\epsilon_c,
\end{equation}
using small horizon cutoffs and a radial mesh refined near $x=1$ and $x=x_c$. The additive constant in $r_*$ is immaterial; it only changes the phases of the extracted amplitudes.

For positive frequency the second-order master equation is evolved as the complex first-order system
\begin{equation}
\label{eq:numerical_first_order_system}
 \frac{\dd}{\dd r_*}
 \begin{pmatrix}
 \Psi\\ \Pi
 \end{pmatrix}
 =
 \begin{pmatrix}
 \Pi\\ -[\omega^2-V(r)]\Psi
 \end{pmatrix},
 \qquad \Pi=\frac{\dd\Psi}{\dd r_*}.
\end{equation}
The integration starts close to the black-hole horizon with the purely ingoing normalization
\begin{equation}
\label{eq:numerical_left_initial_data}
 \Psi(r_{*,L})=e^{-i\omega r_{*,L}},
 \qquad
 \Pi(r_{*,L})=-i\omega e^{-i\omega r_{*,L}},
\end{equation}
and is carried to a point close to the cosmological horizon. In the final asymptotic segment the solution is projected onto the two plane waves in Eq.~\eqref{eq:boundary_conditions} by
\begin{align}
\label{eq:numerical_amplitude_projection}
 A_{\rm in}(r_*)&=\frac{1}{2}e^{+i\omega r_*}
 \left(\Psi+\frac{i}{\omega}\Pi\right),\\
 A_{\rm out}(r_*)&=\frac{1}{2}e^{-i\omega r_*}
 \left(\Psi-\frac{i}{\omega}\Pi\right).
\end{align}
The quoted amplitudes are the mean values of these two nearly constant quantities over the near-cosmological-horizon tail. As a numerical diagnostic we monitor
\begin{equation}
\label{eq:numerical_flux_error}
 \Delta_{\rm flux}=|\Gamma+\mathcal{R}-1|,
\end{equation}
whose maximum over the displayed data set is $6.9\times10^{-5}$. At exactly zero frequency the scalar $\ell=0$ de Sitter mode is treated separately by matching the regular solution
\begin{equation}
\label{eq:numerical_scalar_zero_solution}
 \phi_0(x)=C_1+C_2\int^x\frac{\dd y}{y^2 f(y)}
\end{equation}
to the two horizon regions. This gives the analytic limit
\begin{equation}
\label{eq:numerical_scalar_zero_limit}
 \Gamma^{(s)}_0(0)=
 \frac{4r_b^2r_c^2}{(r_b^2+r_c^2)^2}
 =\frac{4x_c^2}{(1+x_c^2)^2}
 \quad (r_b=1),
\end{equation}
which explains the nonzero scalar intercept shown in Fig.~\ref{fig:ehds_lambda_greybody}. All nonzero-frequency points in the greybody plots and all emission integrals use the direct positive-frequency integrations described above, while WKB is used only for the comparison below.

For comparison we also evaluate a WKB approximation at the peak of each effective potential. The first-order barrier formula is
\begin{equation}
\label{eq:first_order_wkb}
 \Gamma_{\rm WKB}^{(1)}(\omega)
 =\left[1+\exp\left(\frac{2\pi(V_0-\omega^2)}{\sqrt{-2V_0''}}\right)\right]^{-1},
\end{equation}
Here $V_0$ is the value of the effective potential at its maximum, $V_0''=\dd^2V/\dd r_*^2$ is evaluated at the same point and is negative at the barrier top, and primes denote derivatives with respect to the tortoise coordinate. The sixth-order calculation uses the standard Iyer--Will higher-order expansion and its greybody-factor continuation \cite{IyerWill1987,Konoplya2003WKB,KonoplyaZhidenkoZinhailo2019}. Operationally, the derivatives $V_0^{(m)}=\dd^mV/\dd r_*^m$ through $m=12$ are evaluated at the peak, the sixth-order algebraic condition is solved for the real barrier parameter $\nu(\omega)$, and the transmission probability is reconstructed as
\begin{equation}
\label{eq:sixth_order_wkb_gamma}
 \Gamma_{\rm WKB}^{(6)}(\omega)=\left[1+e^{2\pi\nu(\omega)}\right]^{-1}.
\end{equation}
Here $\nu(\omega)$ is the sixth-order WKB barrier parameter obtained from the peak derivatives $V_0^{(m)}$. The WKB method has been extensively applied across a number of spectral problems and finding grey-body factors of various compact objects \cite{Bolokhov:2026kqu,Konoplya:2010vz,Lutfuoglu:2025ljm,Skvortsova:2026jtx,Breton:2017hwe,Bolokhov:2024bke,Konoplya:2024hfg,Fernando:2016ftj,Konoplya:2009hv,Bolokhov:2025lnt,Karmakar:2023cwg,Skvortsova:2024eqi,Konoplya:2023moy,Guo:2020caw,Bolokhov:2026uol,Kodama:2009bf,Konoplya:2019ppy,Skvortsova:2024wly,Konoplya:2024kih,Bolokhov:2026dzn,Konoplya:2025hgp,Malik:2026lfj,Eniceicu:2019npi,Lutfuoglu:2025hjy,Bolokhov:2025egl,Bolokhov:2023dxq,Konoplya:2010kv,Lutfuoglu:2025ljm,Bolokhov:2023bwm,Bolokhov:2023ruj,Skvortsova:2025cah,Skvortsova:2024atk,Konoplya:2023ahd,Arbelaez:2026eaz}.
The WKB approximation is used here only as a local check near the top of the barrier. All emission rates reported below use the direct numerical greybody factors.

\section{Greybody factors}
\label{sec:greybody_results}

\begin{figure*}[t]
\centering
\includegraphics[width=\textwidth]{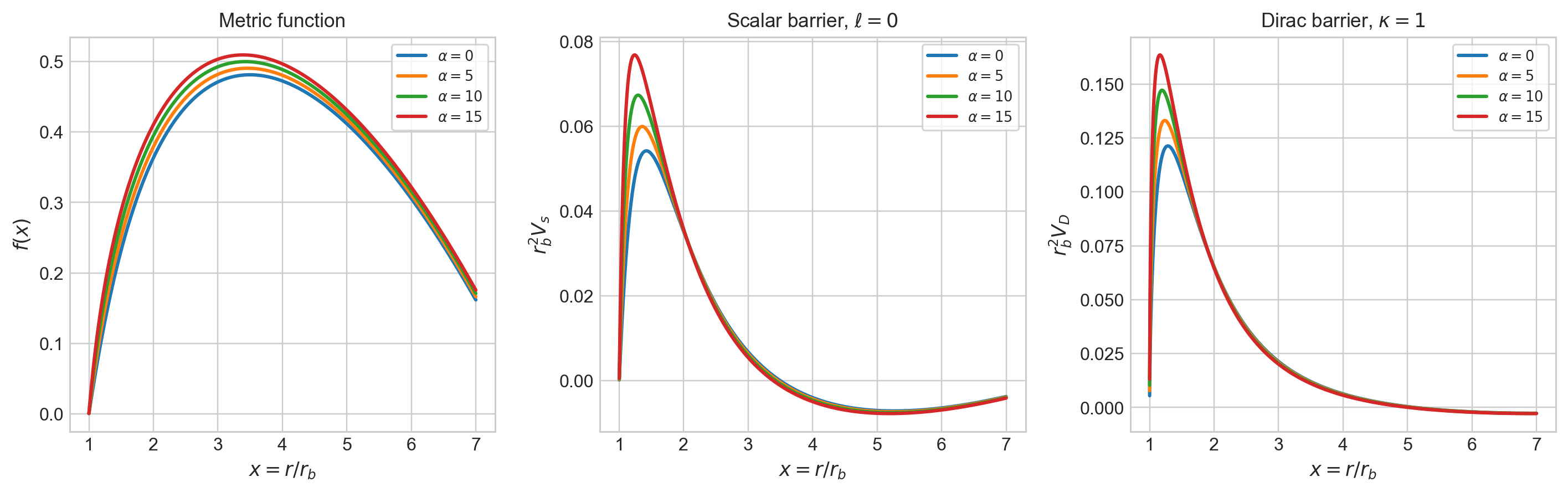}
\caption{Metric function and representative effective potentials for the reference family $q=Q/r_b=0.60$, $\lambda=\Lambda r_b^2=0.04$ and $\alpha=a/r_b^2=0,5,10,15$. The scalar panel uses the dominant $\ell=0$ mode, and the Dirac panel uses $\kappa=1$. Increasing $\alpha$ raises the barriers and moves their maxima closer to the event horizon.}
\label{fig:ehds_metric_potentials}
\end{figure*}

\begin{figure*}[t]
\centering
\includegraphics[width=\textwidth]{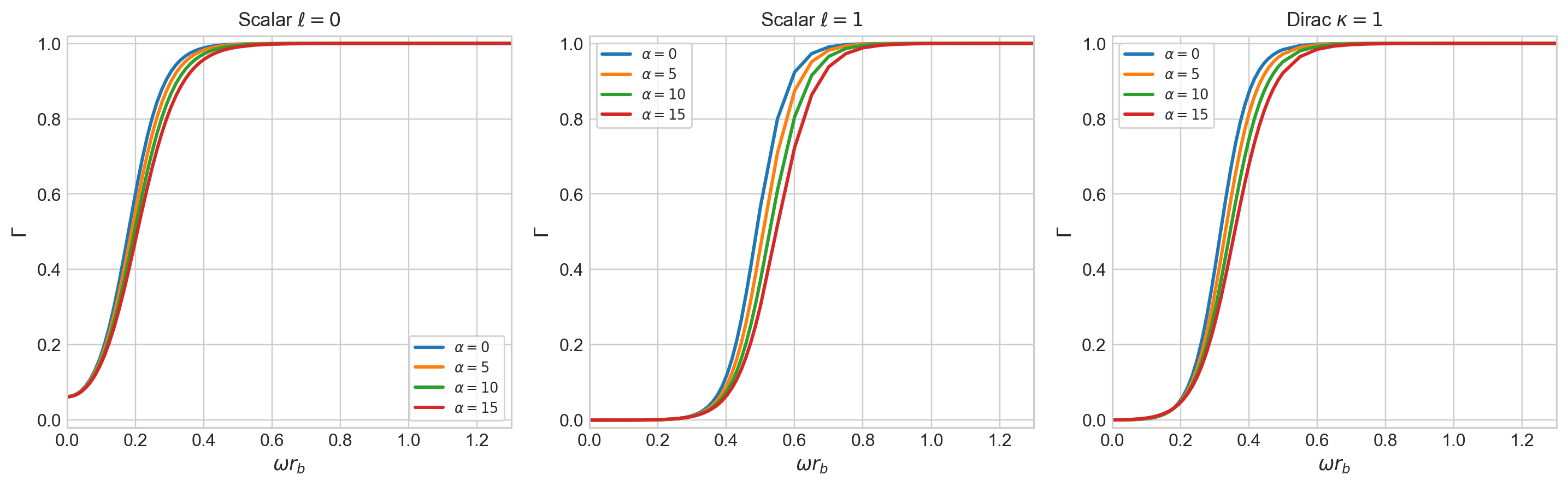}
\caption{Direct numerical greybody factors for neutral scalar and Dirac test fields for $q=Q/r_b=0.60$, $\lambda=\Lambda r_b^2=0.04$ and $\alpha=a/r_b^2=0,5,10,15$. The finite de Sitter scattering region is $1<x<x_c$. The horizontal range is cut at $\omega r_b=1.3$, where the displayed channels are already close to full transmission. Larger $\alpha$ makes the dominant scalar and Dirac modes less transparent at a given frequency near the threshold.}
\label{fig:ehds_greybody}
\end{figure*}

Figure~\ref{fig:ehds_metric_potentials} shows the lapse function and two representative potentials. Increasing $\alpha$ at fixed $q$ and $\lambda$ slightly enlarges the cosmological horizon, moves the dominant barrier peaks inward, and raises the peak heights. This is reflected directly in the transmission curves of Fig.~\ref{fig:ehds_greybody}: the turn-on of the scalar and Dirac greybody factors shifts to larger frequency as the Euler--Heisenberg correction grows.

\begin{figure*}[t]
\centering
\includegraphics[width=0.92\textwidth]{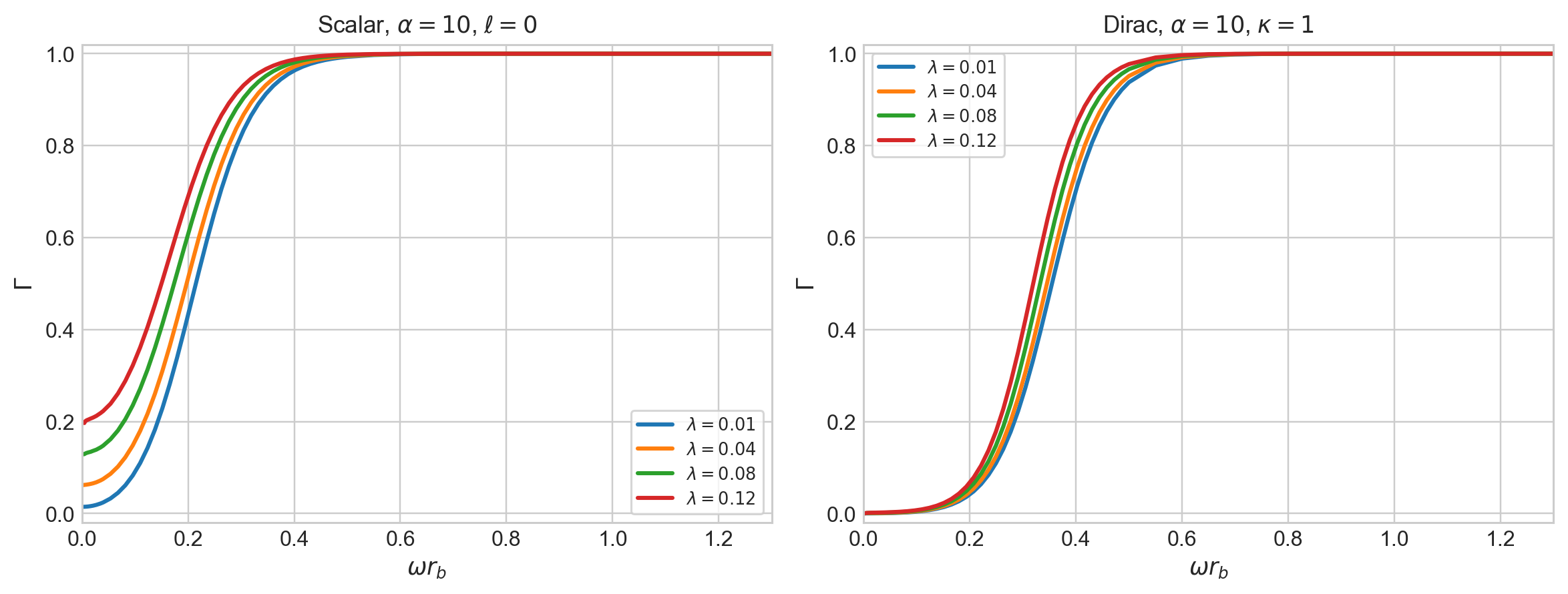}
\caption{Greybody factors for representative neutral scalar and Dirac channels at fixed $q=Q/r_b=0.60$ and $\alpha=a/r_b^2=10$, with $\lambda=\Lambda r_b^2=0.01,0.04,0.08,0.12$. Positive-frequency points are direct numerical integrations, not WKB estimates. The scalar $\ell=0$ point at $\omega=0$ shows the analytic de Sitter zero-frequency limit $\Gamma_0^{(s)}(0)=4x_c^2/(1+x_c^2)^2$; the ultra-low-frequency numerical extraction below $\omega r_b=4\times10^{-3}$ is omitted from that panel because the incoming/outgoing amplitudes become ill-conditioned there. The horizontal range is cut at $\omega r_b=1.3$ because all curves are already close to full transmission. Increasing $\lambda$ lowers the threshold region of both displayed channels.}
\label{fig:ehds_lambda_greybody}
\end{figure*}

Figure~\ref{fig:ehds_lambda_greybody} isolates the complementary role of the cosmological constant by holding $q=0.60$ and $\alpha=10$ fixed while varying $\lambda=\Lambda r_b^2$. The curves are obtained from the same direct radial integrations used elsewhere; the only special treatment is the scalar $\ell=0$ endpoint at the origin. In a two-horizon de Sitter static patch the minimally coupled scalar s-wave has a finite zero-frequency limit, obtained by matching the $\omega\to0$ solution $\phi=C_1+C_2\int dx/(x^2f)$ to horizon waves, namely $\Gamma_0^{(s)}(0)=4x_c^2/(1+x_c^2)^2$ for $r_b=1$. This is why the scalar panel need not vanish at $\omega=0$; it is not a WKB artifact. The Dirac channel still tends to zero. Increasing $\lambda$ moves the cosmological horizon inward, from $x_c=16.64656$ at $\lambda=0.01$ to $x_c=4.25742$ at $\lambda=0.12$, and lowers the turn-on frequencies of the dominant greybody channels. The scalar $\ell=0$ half-transmission point decreases from $\omega r_b=0.21474$ to $0.14993$, while the Dirac $\kappa=1$ point decreases from $0.35480$ to $0.31940$. Thus, along this fixed-$\alpha$ slice, a larger positive cosmological constant makes the finite static patch more transparent near threshold. This trend is opposite to the fixed-$\lambda$ increase of $\alpha$, which raises the barriers and shifts the greybody onset upward. At larger frequencies the curves merge as $\Gamma\to1$, so the visible effect of $\lambda$ is concentrated in the low- and intermediate-frequency part of the spectrum.

Table~\ref{tab:ehds_half_points} quantifies the shift. From $\alpha=0$ to $\alpha=15$, the half-transmission frequency increases from $0.17986$ to $0.20571$ for the scalar $\ell=0$ mode, from $0.48829$ to $0.54592$ for the scalar $\ell=1$ mode, and from $0.31747$ to $0.35870$ for the Dirac $\kappa=1$ mode. Thus, in this chosen branch the Euler--Heisenberg correction makes the exterior barrier more opaque, rather than more transparent.

\begin{table*}[t]
\centering
\caption{Barrier data and direct numerical half-transmission frequencies for representative scalar and Dirac modes with $q=Q/r_b=0.60$, $\lambda=\Lambda r_b^2=0.04$ and $\alpha=a/r_b^2=0,5,10,15$. The peak position and peak height are measured in units of $r_b$ and $r_b^{-2}$, respectively.}
\label{tab:ehds_half_points}
\begingroup
\small
\setlength{\tabcolsep}{4.5pt}
\renewcommand{\arraystretch}{1.03}
\begin{ruledtabular}
\begin{tabular}{ccccc}
$\alpha$ & mode & $x_{\rm peak}$ & $r_b^2V_0$ & $\omega_{1/2}r_b$ \\
$0$ & scalar $\ell=0$ & $1.42061$ & $5.419\times10^{-2}$ & $0.17986$ \\
$0$ & scalar $\ell=1$ & $1.51811$ & $2.600\times10^{-1}$ & $0.48829$ \\
$0$ & Dirac $\kappa=1$ & $1.28524$ & $1.213\times10^{-1}$ & $0.31747$ \\
$5$ & scalar $\ell=0$ & $1.35675$ & $5.994\times10^{-2}$ & $0.18867$ \\
$5$ & scalar $\ell=1$ & $1.46808$ & $2.808\times10^{-1}$ & $0.50802$ \\
$5$ & Dirac $\kappa=1$ & $1.23727$ & $1.331\times10^{-1}$ & $0.33150$ \\
$10$ & scalar $\ell=0$ & $1.29114$ & $6.732\times10^{-2}$ & $0.19731$ \\
$10$ & scalar $\ell=1$ & $1.42109$ & $3.035\times10^{-1}$ & $0.52695$ \\
$10$ & Dirac $\kappa=1$ & $1.19337$ & $1.472\times10^{-1}$ & $0.34533$ \\
$15$ & scalar $\ell=0$ & $1.23727$ & $7.680\times10^{-2}$ & $0.20571$ \\
$15$ & scalar $\ell=1$ & $1.38057$ & $3.283\times10^{-1}$ & $0.54592$ \\
$15$ & Dirac $\kappa=1$ & $1.16084$ & $1.635\times10^{-1}$ & $0.35870$ \\
\end{tabular}
\end{ruledtabular}
\endgroup
\end{table*}

\begin{figure*}[t]
\centering
\includegraphics[width=0.92\textwidth]{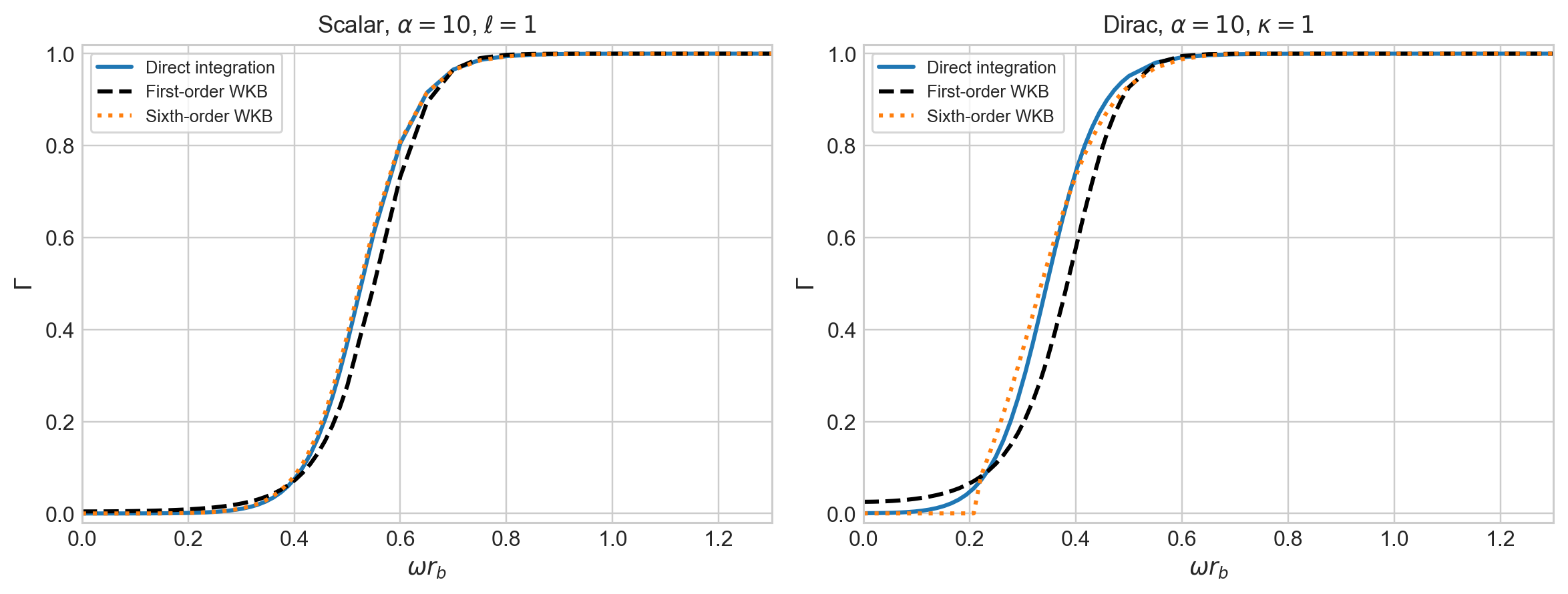}
\caption{Direct transmission coefficients compared with first- and sixth-order WKB estimates for representative modes at $q=Q/r_b=0.60$, $\lambda=\Lambda r_b^2=0.04$ and $\alpha=a/r_b^2=10$. The horizontal range is cut at $\omega r_b=1.3$, where the plotted transmissions are already close to unity, so that the differences between curves near the barrier turn-on remain visible. The sixth-order WKB approximation tracks the numerical barrier-top transition much more closely than the first-order formula.}
\label{fig:ehds_wkb}
\end{figure*}

The WKB comparison is shown in Fig.~\ref{fig:ehds_wkb} and Tables~\ref{tab:ehds_wkb_samples}--\ref{tab:ehds_wkb_half}. The first-order WKB curve already captures the qualitative barrier turn-on, but it underestimates the transmission near the representative half-transmission region. The sixth-order correction is substantially better. For the scalar $\ell=1$ mode at $\alpha=10$, the half-transmission error improves from $4.56\%$ at first order to $-0.58\%$ at sixth order. For the Dirac $\kappa=1$ mode the corresponding error improves from $11.07\%$ to $-2.91\%$. Pointwise errors away from the exact half-transmission point are larger, especially for the Dirac modes below the peak, which is why the direct integration is retained for all quantitative spectra. The added $\ell=2$ and $\kappa=2$ rows show the expected multipole trend: sixth-order WKB is very accurate for the higher scalar multipole and becomes noticeably better for the higher Dirac multipole, whereas the lower modes, especially the Dirac $\kappa=1$ channel below the peak, remain insufficiently described by WKB for precision greybody and emission calculations.

\begin{table*}[t]
\centering
\caption{Pointwise WKB comparison near the numerical half-transmission region for $q=Q/r_b=0.60$, $\lambda=\Lambda r_b^2=0.04$ and $\alpha=a/r_b^2=10$, including the next higher scalar and Dirac multipoles. Errors are relative to the direct integration result.}
\label{tab:ehds_wkb_samples}
\begin{ruledtabular}
\begin{tabular}{cccccccc}
field & mode & $\omega r_b$ & $\Gamma_{\rm num}$ & $\Gamma_{\rm WKB}^{(1)}$ & error & $\Gamma_{\rm WKB}^{(6)}$ & error \\
Scalar & $\ell=1$ & $0.5000$ & $0.37079$ & $0.27745$ & $-25.17$\% & $0.38769$ & $4.56$\% \\
Scalar & $\ell=1$ & $0.5500$ & $0.61049$ & $0.49540$ & $-18.85$\% & $0.62274$ & $2.01$\% \\
Scalar & $\ell=1$ & $0.6000$ & $0.80503$ & $0.73297$ & $-8.95$\% & $0.80938$ & $0.54$\% \\
Dirac & $\kappa=1$ & $0.3327$ & $0.43769$ & $0.28826$ & $-34.14$\% & $0.48979$ & $11.90$\% \\
Dirac & $\kappa=1$ & $0.3467$ & $0.50661$ & $0.33870$ & $-33.14$\% & $0.54579$ & $7.73$\% \\
Dirac & $\kappa=1$ & $0.3606$ & $0.57482$ & $0.39540$ & $-31.21$\% & $0.59979$ & $4.34$\% \\
Scalar & $\ell=2$ & $0.8000$ & $0.19352$ & $0.16498$ & $-14.75$\% & $0.19354$ & $0.01$\% \\
Scalar & $\ell=2$ & $0.8500$ & $0.39658$ & $0.33507$ & $-15.51$\% & $0.39583$ & $-0.19$\% \\
Scalar & $\ell=2$ & $0.9000$ & $0.64098$ & $0.57631$ & $-10.09$\% & $0.63996$ & $-0.16$\% \\
Dirac & $\kappa=2$ & $0.6500$ & $0.29411$ & $0.23180$ & $-21.18$\% & $0.27507$ & $-6.47$\% \\
Dirac & $\kappa=2$ & $0.7000$ & $0.53432$ & $0.43917$ & $-17.81$\% & $0.51527$ & $-3.57$\% \\
Dirac & $\kappa=2$ & $0.7500$ & $0.75750$ & $0.68562$ & $-9.49$\% & $0.75133$ & $-0.81$\% \\
\end{tabular}
\end{ruledtabular}
\end{table*}

\begin{table*}[t]
\centering
\caption{Half-transmission comparison for the same WKB comparison at $q=Q/r_b=0.60$, $\lambda=\Lambda r_b^2=0.04$ and $\alpha=a/r_b^2=10$. The sixth-order approximation gives percent-level agreement with the direct result at the barrier top.}
\label{tab:ehds_wkb_half}
\begin{ruledtabular}
\begin{tabular}{cccccccc}
field & mode & $x_{\rm peak}$ & $\omega_{1/2}^{\rm num}r_b$ & $\omega_{1/2}^{(1)}r_b$ & error & $\omega_{1/2}^{(6)}r_b$ & error \\
Scalar & $\ell=1$ & $1.42179$ & $0.52695$ & $0.55097$ & $4.56$\% & $0.52389$ & $-0.58$\% \\
Dirac & $\kappa=1$ & $1.19330$ & $0.34533$ & $0.38356$ & $11.07$\% & $0.33527$ & $-2.91$\% \\
\end{tabular}
\end{ruledtabular}
\end{table*}

\clearpage

\section{Hawking emission of neutral particles}
\label{sec:emission_rates}

The black-hole contribution to the spectral energy flux of a neutral massless scalar field is computed as
\begin{equation}
\label{eq:scalar_emission}
 \frac{\dd^2E_s}{\dd t\,\dd\omega}
 =\frac{1}{2\pi}\sum_{\ell=0}^{\infty}(2\ell+1)
 \frac{\Gamma_\ell(\omega)\,\omega}{\exp(\omega/T)-1}.
\end{equation}
Here $\dd^2E_s/(\dd t\,\dd\omega)$ is the scalar spectral power, $\Gamma_\ell$ is the scalar greybody factor, $T$ is the chosen black-hole or effective temperature, and $2\ell+1$ is the scalar angular degeneracy.
For one neutral four-component massless Dirac field we use
\begin{equation}
\label{eq:dirac_emission}
 \frac{\dd^2E_D}{\dd t\,\dd\omega}
 =\frac{1}{2\pi}\sum_{\kappa=1}^{\infty}4\kappa
 \frac{\Gamma_\kappa(\omega)\,\omega}{\exp(\omega/T)+1}.
\end{equation}
Here $\dd^2E_D/(\dd t\,\dd\omega)$ is the Dirac spectral power, $\Gamma_\kappa$ is the Dirac greybody factor, $4\kappa$ is the degeneracy of a neutral four-component massless Dirac channel, and the plus sign in the denominator is the Fermi--Dirac factor.
The sums are truncated at $\ell=5$ and $\kappa=5$. At the spectral peaks the omitted channels are negligible for the chosen data set: the largest last-channel contribution is below $2.3\times10^{-16}$ for the scalar spectra and below $5.5\times10^{-13}$ for the Dirac spectra. The integrated powers are
\begin{equation}
\label{eq:integrated_powers}
 P_s=\int_0^\infty\frac{\dd^2E_s}{\dd t\,\dd\omega}\,\dd\omega,
 \qquad
 P_D=\int_0^\infty\frac{\dd^2E_D}{\dd t\,\dd\omega}\,\dd\omega .
\end{equation}
Here $P_s$ and $P_D$ are the frequency-integrated scalar and Dirac powers, respectively.
All entries below are dimensionless $r_b^2P$ values.

\begin{figure*}[t]
\centering
\includegraphics[width=0.72\textwidth]{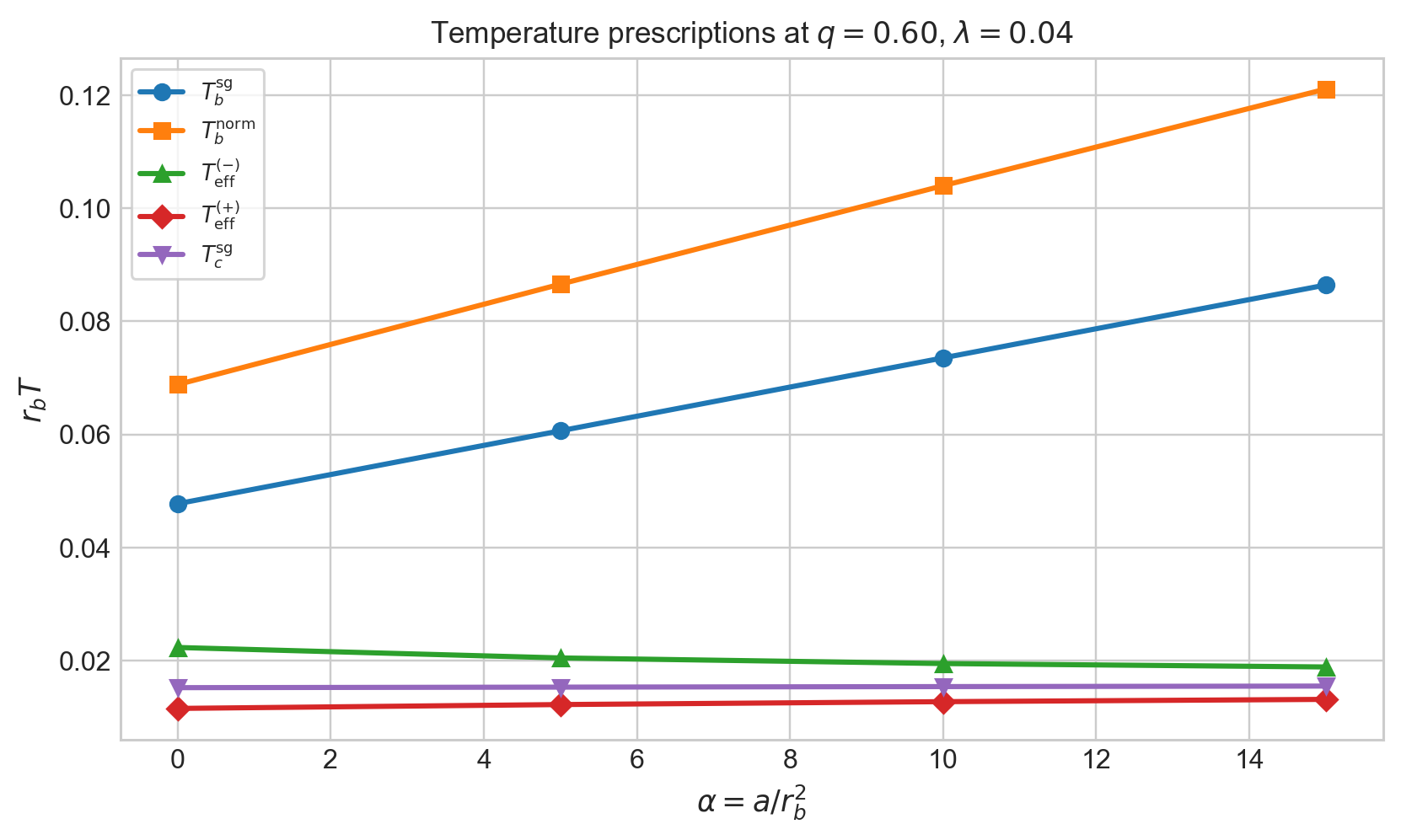}
\caption{Representative temperature prescriptions for the reference de Sitter static patch with $q=Q/r_b=0.60$, $\lambda=\Lambda r_b^2=0.04$ and $\alpha=a/r_b^2=0,5,10,15$. The event-horizon surface-gravity and Bousso--Hawking-normalized temperatures increase with $\alpha$, while the cosmological-horizon temperature changes only weakly. Effective static-patch temperatures are therefore much lower than the event-horizon values.}
\label{fig:ehds_temperature_choices}
\end{figure*}

In de Sitter space the choice of $T$ in Eqs.~\eqref{eq:scalar_emission}--\eqref{eq:dirac_emission} is part of the physical prescription. Figure~\ref{fig:ehds_temperature_choices} compares several possible temperatures for the same backgrounds, and Table~\ref{tab:ehds_power} gives the corresponding integrated powers. The bare event-horizon temperature and the Bousso--Hawking-normalized temperature both increase with $\alpha$. Although the greybody factors become more suppressed, this temperature increase dominates the integrated flux. The total scalar-plus-Dirac power rises by a factor of $8.88$ between $\alpha=0$ and $\alpha=15$ when $T=T_b^{\rm sg}$, and by a factor of $8.05$ when $T=T_b^{\rm norm}$.

\begin{figure*}[t]
\centering
\includegraphics[width=\textwidth]{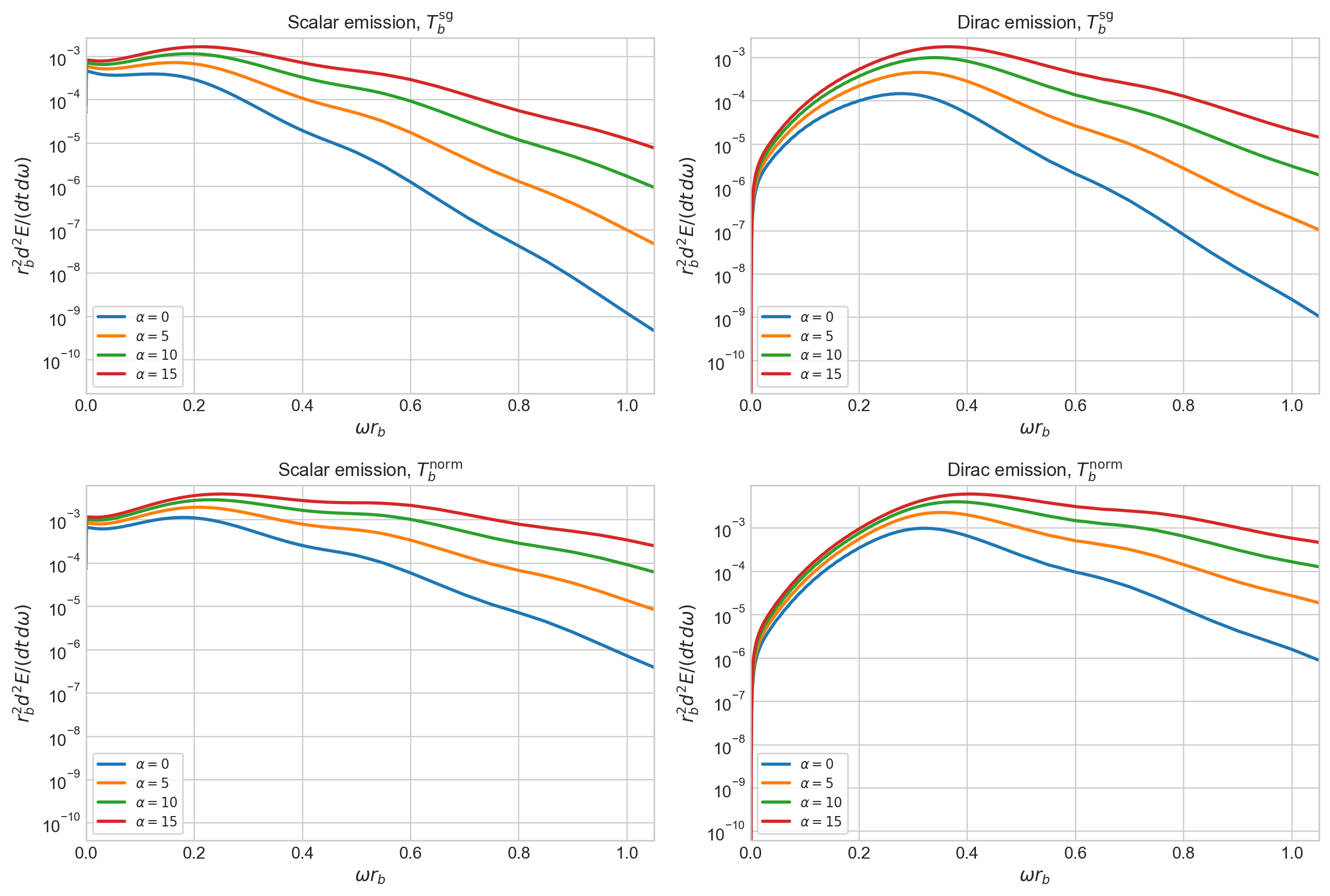}
\caption{Neutral scalar and Dirac emission spectra for $q=Q/r_b=0.60$, $\lambda=\Lambda r_b^2=0.04$ and $\alpha=a/r_b^2=0,5,10,15$, using the event-horizon temperatures $T_b^{\rm sg}$ and $T_b^{\rm norm}$. The same direct numerical greybody factors are used in both rows. The frequency range is truncated to the peak region, $0<\omega r_b\leq1.05$, and the curves are sampled on a denser plotting grid so that the spectral maxima are easier to resolve. The Bousso--Hawking normalization raises the temperature and therefore the luminosity relative to the bare surface-gravity prescription.}
\label{fig:ehds_emission}
\end{figure*}

\begin{table*}[t]
\centering
\caption{Integrated neutral-particle emission rates for $q=Q/r_b=0.60$, $\lambda=\Lambda r_b^2=0.04$ and $\alpha=a/r_b^2=0,5,10,15$, using the temperature prescriptions listed in the second column. The scalar column is for one massless scalar field; the Dirac column is for one neutral four-component massless Dirac field. The powers are dimensionless $r_b^2P$.}
\label{tab:ehds_power}
\begin{ruledtabular}
\begin{tabular}{cccccc}
$\alpha$ & temperature prescription & $r_bT$ & $r_b^2P_s$ & $r_b^2P_D$ & $r_b^2(P_s+P_D)$ \\
$0$ & $T_b^{\rm sg}$ & $4.775\times10^{-2}$ & $9.969\times10^{-5}$ & $3.301\times10^{-5}$ & $1.327\times10^{-4}$ \\
$0$ & $T_b^{\rm norm}$ & $6.882\times10^{-2}$ & $3.344\times10^{-4}$ & $2.411\times10^{-4}$ & $5.755\times10^{-4}$ \\
$0$ & $T_{\rm eff}^{(-)}$ & $2.231\times10^{-2}$ & $1.079\times10^{-5}$ & $4.134\times10^{-7}$ & $1.120\times10^{-5}$ \\
$0$ & $T_{\rm eff}^{(+)}$ & $1.153\times10^{-2}$ & $2.305\times10^{-6}$ & $3.483\times10^{-8}$ & $2.340\times10^{-6}$ \\
$5$ & $T_b^{\rm sg}$ & $6.064\times10^{-2}$ & $2.063\times10^{-4}$ & $1.124\times10^{-4}$ & $3.187\times10^{-4}$ \\
$5$ & $T_b^{\rm norm}$ & $8.657\times10^{-2}$ & $6.959\times10^{-4}$ & $6.683\times10^{-4}$ & $1.364\times10^{-3}$ \\
$5$ & $T_{\rm eff}^{(-)}$ & $2.046\times10^{-2}$ & $8.495\times10^{-6}$ & $3.281\times10^{-7}$ & $8.823\times10^{-6}$ \\
$5$ & $T_{\rm eff}^{(+)}$ & $1.222\times10^{-2}$ & $2.588\times10^{-6}$ & $5.519\times10^{-8}$ & $2.644\times10^{-6}$ \\
$10$ & $T_b^{\rm sg}$ & $7.353\times10^{-2}$ & $3.738\times10^{-4}$ & $2.769\times10^{-4}$ & $6.507\times10^{-4}$ \\
$10$ & $T_b^{\rm norm}$ & $1.040\times10^{-1}$ & $1.259\times10^{-3}$ & $1.421\times10^{-3}$ & $2.679\times10^{-3}$ \\
$10$ & $T_{\rm eff}^{(-)}$ & $1.947\times10^{-2}$ & $7.404\times10^{-6}$ & $3.213\times10^{-7}$ & $7.726\times10^{-6}$ \\
$10$ & $T_{\rm eff}^{(+)}$ & $1.273\times10^{-2}$ & $2.806\times10^{-6}$ & $7.925\times10^{-8}$ & $2.885\times10^{-6}$ \\
$15$ & $T_b^{\rm sg}$ & $8.642\times10^{-2}$ & $6.186\times10^{-4}$ & $5.596\times10^{-4}$ & $1.178\times10^{-3}$ \\
$15$ & $T_b^{\rm norm}$ & $1.211\times10^{-1}$ & $2.064\times10^{-3}$ & $2.571\times10^{-3}$ & $4.635\times10^{-3}$ \\
$15$ & $T_{\rm eff}^{(-)}$ & $1.887\times10^{-2}$ & $6.779\times10^{-6}$ & $3.390\times10^{-7}$ & $7.118\times10^{-6}$ \\
$15$ & $T_{\rm eff}^{(+)}$ & $1.313\times10^{-2}$ & $2.981\times10^{-6}$ & $1.065\times10^{-7}$ & $3.087\times10^{-6}$ \\
\end{tabular}
\end{ruledtabular}
\end{table*}

\begin{figure*}[t]
\centering
\includegraphics[width=0.72\textwidth]{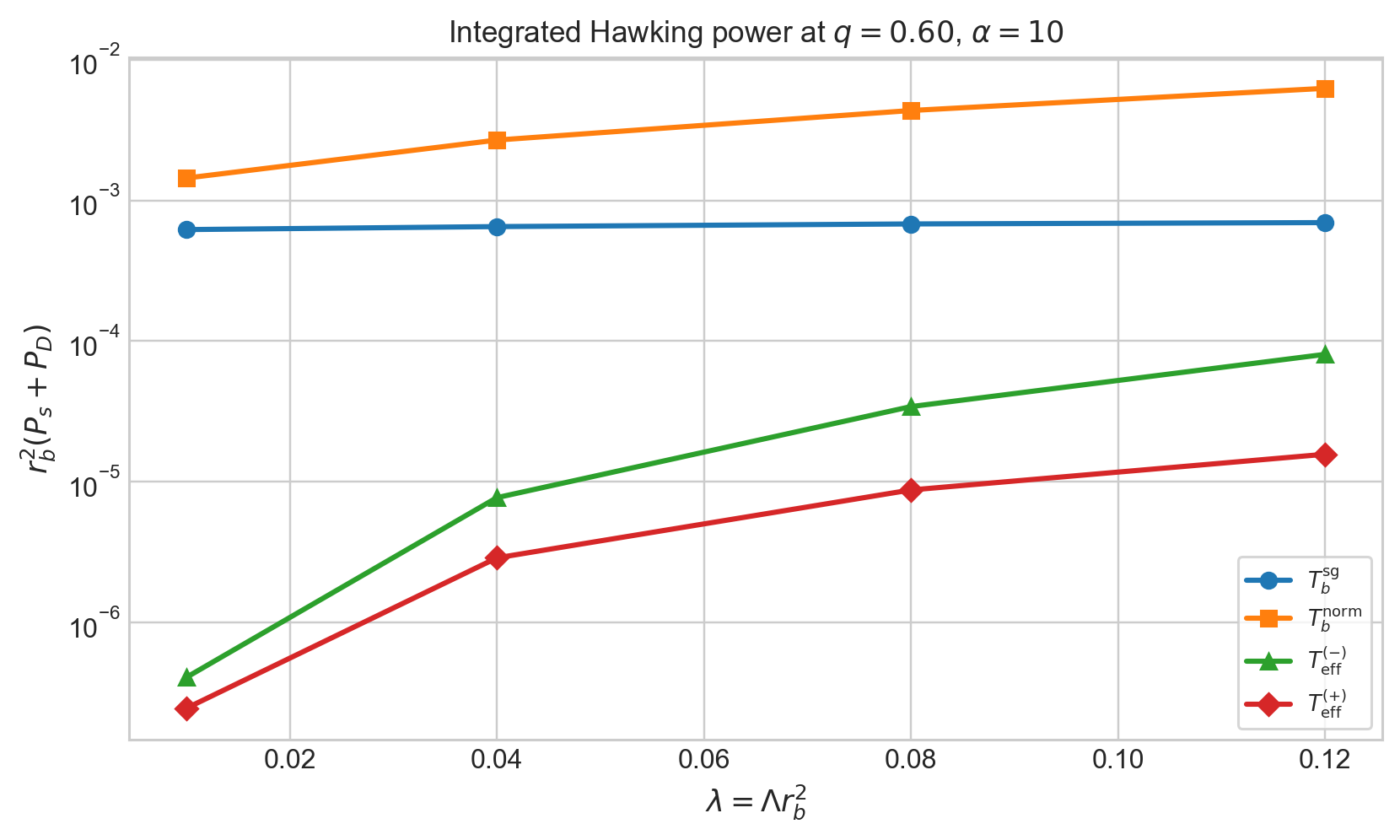}
\caption{Integrated neutral scalar-plus-Dirac power as a function of $\lambda=\Lambda r_b^2$ at fixed $q=0.60$ and $\alpha=10$. The powers are evaluated with the same direct numerical greybody factors and with the temperature prescriptions shown in the legend.}
\label{fig:ehds_lambda_power_trend}
\end{figure*}

To isolate the role of the cosmological constant, we also repeated the Hawking-power calculation at fixed $q=0.60$ and fixed $\alpha=10$ while varying $\lambda=\Lambda r_b^2$. Table~\ref{tab:ehds_lambda_power} and Fig.~\ref{fig:ehds_lambda_power_trend} show that increasing $\lambda$ brings the cosmological horizon inward, from $x_c=16.64656$ at $\lambda=0.01$ to $x_c=4.25742$ at $\lambda=0.12$. The direct greybody calculation gives a lower dominant barrier over this interval: the half-transmission frequency changes from $\omega r_b=0.21474$ to $0.14993$ for the scalar $\ell=0$ channel and from $0.35480$ to $0.31940$ for the Dirac $\kappa=1$ channel. Consequently, the bare-temperature power rises modestly even though $T_b^{\rm sg}$ itself decreases. The Bousso--Hawking-normalized and effective static-patch prescriptions respond more strongly because their assigned temperatures grow as the static patch contracts.

\begin{table*}[t]
\centering
\caption{Fixed-$q$, fixed-$\alpha$ cosmological-constant scan of the neutral scalar-plus-Dirac Hawking power. Here $q=0.60$, $\alpha=10$, and all powers are dimensionless $r_b^2P$.}
\label{tab:ehds_lambda_power}
\begin{ruledtabular}
\begin{tabular}{ccccccc}
$\lambda$ & $x_c$ & $r_bT_b^{\rm sg}$ & $r_b^2P_{\rm sg}$ & $r_b^2P_{\rm norm}$ & $r_b^2P_{\rm eff}^{(-)}$ & $r_b^2P_{\rm eff}^{(+)}$ \\
$0.01$ & $16.64656$ & $7.592\times10^{-2}$ & $6.196\times10^{-4}$ & $1.439\times10^{-3}$ & $4.084\times10^{-7}$ & $2.465\times10^{-7}$ \\
$0.04$ & $7.95908$ & $7.353\times10^{-2}$ & $6.507\times10^{-4}$ & $2.679\times10^{-3}$ & $7.726\times10^{-6}$ & $2.885\times10^{-6}$ \\
$0.08$ & $5.39932$ & $7.035\times10^{-2}$ & $6.796\times10^{-4}$ & $4.344\times10^{-3}$ & $3.426\times10^{-5}$ & $8.764\times10^{-6}$ \\
$0.12$ & $4.25742$ & $6.716\times10^{-2}$ & $6.946\times10^{-4}$ & $6.245\times10^{-3}$ & $8.052\times10^{-5}$ & $1.571\times10^{-5}$ \\
\end{tabular}
\end{ruledtabular}
\end{table*}

The effective static-patch temperatures lead to a different conclusion. The temperature $T_{\rm eff}^{(-)}$ is controlled by the relatively cold cosmological horizon and decreases mildly across the reference sequence. With this choice the total power decreases from $1.12\times10^{-5}$ at $\alpha=0$ to $7.12\times10^{-6}$ at $\alpha=15$, despite the increase of the event-horizon temperature. The alternative harmonic choice $T_{\rm eff}^{(+)}$ is lower still, but rises slightly with $\alpha$ and gives a modest power increase. Thus the temperature ambiguity is not a small normalization issue: it can change both the magnitude and the qualitative trend of the luminosity.

The spectra in Fig.~\ref{fig:ehds_emission} should be interpreted as black-hole-horizon emission into the static patch. A full nonequilibrium treatment would also include radiation associated with the cosmological horizon and, for charged emitted quanta, electrostatic chemical potentials. The present calculation deliberately avoids those additional ingredients by considering neutral scalar and neutral Dirac test fields and by displaying the dependence on the chosen temperature explicitly.

\section{Conclusion}
\label{sec:conclusion}

We have computed neutral scalar and neutral massless Dirac transmission coefficients and Hawking spectra for the positive-cosmological-constant branch of the Einstein--Euler--Heisenberg black hole. The calculation keeps the event horizon fixed, treats the finite static patch between the event and cosmological horizons as a two-sided scattering region, and obtains the quoted greybody factors and emission rates from direct integration of the radial master equations. The benchmark sequence fixes $q=Q/r_b=0.60$ and $\lambda=\Lambda r_b^2=0.04$ while varying the dimensionless nonlinear-electrodynamic coupling $\alpha=a/r_b^2$.

The main scattering result is that the Euler--Heisenberg correction makes this reference static patch less transparent. From $\alpha=0$ to $\alpha=15$, the dominant half-transmission frequencies increase from $0.17986$ to $0.20571$ for the scalar $\ell=0$ channel, from $0.48829$ to $0.54592$ for the scalar $\ell=1$ channel, and from $0.31747$ to $0.35870$ for the Dirac $\kappa=1$ channel. A sixth-order WKB calculation gives an independent barrier-top check and reduces the first-order half-transmission mismatch to the percent level for the representative modes, but the direct integrations remain necessary for the lowest channels that dominate the Hawking flux. At fixed $q=0.60$ and $\alpha=10$, increasing the positive cosmological constant has the opposite greybody effect: the cosmological horizon moves inward, the dominant thresholds decrease, and the finite static patch becomes more transparent near onset.

The emission result depends crucially on the temperature prescription. With the black-hole event-horizon surface-gravity temperature or the Bousso--Hawking-normalized temperature, the increasing temperature overcomes the additional greybody suppression and the total neutral scalar-plus-Dirac luminosity grows with $\alpha$, by factors of $8.88$ and $8.05$, respectively, across the reference sequence. With the effective temperature $T_{\rm eff}^{(-)}$, the rates are much smaller and the total power decreases from $1.12\times10^{-5}$ to $7.12\times10^{-6}$; the harmonic effective prescription $T_{\rm eff}^{(+)}$ remains small but rises mildly. This sensitivity is a characteristic feature of de Sitter black-hole evaporation and must be kept explicit in any phenomenological interpretation. The fixed-$q$, fixed-$\alpha$ scan further shows that increasing $\lambda$ can enhance the luminosity by contracting the static patch, lowering the relevant greybody thresholds and amplifying the normalized or effective temperature prescriptions.

The present work is therefore a fixed-background baseline for neutral test fields, not a complete nonequilibrium evaporation model. Natural extensions include charged scalar and Dirac emission with chemical potentials, electromagnetic perturbations in the nonlinear-electrodynamic optical metric, gravitational perturbations, and a treatment in which black-hole and cosmological-horizon fluxes are handled simultaneously.

Notice that the grey-body factors obtained in the present work are in correspondence with quasinormal modes, according to the link found in \cite{Konoplya:2024vuj,Konoplya:2024lir}. and tested in consequent studied \cite{Lutfuoglu:2025mqa,Skvortsova:2024msa,Bolokhov:2024otn,Malik:2024cgb,Dubinsky:2024vbn,Han:2026fpn,Han:2025cal}. 

\bibliographystyle{apsrev4-1}
\bibliography{eh_ds_references}

\end{document}